\documentclass[twocolumn]{aastex63}

\usepackage[normalem]{ulem}

\begin{document}

\title{The Space Environment and Atmospheric Joule Heating of the Habitable Zone Exoplanet TOI700-d}

\correspondingauthor{Ofer Cohen}
\email{ofer\_cohen@uml.edu}

\author{Ofer Cohen}
\affiliation{Lowell Center for Space Science and Technology, University of Massachusetts Lowell, 600 Suffolk Street, Lowell, MA 01854, USA}

\author{ C. Garraffo}
\affiliation{Institute for Applied Computational Science, Harvard University, 33 Oxford St., Cambridge, Massachusetts, USA}
\affiliation{Harvard-Smithsonian Center for Astrophysics, 60 Garden St., Cambridge, Massachusetts, USA}

\author{Sofia-Paraskevi Moschou}
\affiliation{Harvard-Smithsonian Center for Astrophysics, 60 Garden St., Cambridge, Massachusetts, USA}

\author{Jeremy J. Drake}
\affiliation{Harvard-Smithsonian Center for Astrophysics, 60 Garden St., Cambridge, Massachusetts, USA}

\author{J.~D. Alvarado-G\'omez}
\affiliation{Leibniz Institute for Astrophysics Potsdam, An der Sternwarte 16, 14482 Potsdam, Germany}

\author{Alex Glocer}
\affiliation{NASA Goddard Space Flight Center, Greenbelt, MD, USA}

\author{Federico Fraschetti}
\affiliation{Dept. of Planetary Sciences-Lunar and Planetary Laboratory, University of Arizona, Tucson, AZ, 85721, USA}

\begin{abstract}

We investigate the space environment conditions near the Earth-size planet TOI~700~d using a set of numerical models for the stellar corona and wind, the planetary magnetosphere, and the planetary ionosphere. We drive our simulations using a scaled-down stellar input and a scaled-up solar input in order to obtain two independent solutions. We find that for the particular parameters used in our study, the stellar wind conditions near the planet are not very extreme --- slightly stronger than that near the Earth in terms of the stellar wind ram pressure and the intensity of the interplanetary magnetic field. Thus, the space environment near TOI700-d may not be extremely harmful to the planetary atmosphere, assuming the planet resembles the Earth. Nevertheless, we stress that the stellar input parameters and the actual planetary parameters are unconstrained, and different parameters may result in a much greater effect on the atmosphere of TOI700-d. Finally, we compare our results to solar wind measurements in the solar system and stress that modest stellar wind conditions may not guarantee atmospheric retention of exoplanets.

\end{abstract}


\section{Introduction} 
\label{sec:intro}

TOI700 (TIC 150428135) is an M2 dwarf star located at 31.1~pc. It is a slow rotator, with a rotation period of $P~=~54$~days, estimated age of $>1.5$~Gyr, and a very low X-ray and EUV activity, with $L_x<2.4 \times 10^{27}$~erg~s$^{-1}$ \citep{TOI700a}. Recently, a three-planet system has been confirmed using data from the Transiting Exoplanet Survey Satellite \citep[TESS,][]{RickerTESS} and the Spitzer InfraRed Array Camera \citep[IRAC,][]{Fazio04} by \cite{TOI700a} and \cite{TOI700b}. The three-planet system is composed of an Earth-size planet (TOI700-b, $R_p=1.01$~R$_\Earth$) orbiting at $0.0637$~au, a slightly larger planet (TOI700-c, $R_p=2.63$~R$_\Earth$) orbiting at $0.0925$~au, and another Earth-size planet (TOI700-d, $R_p=1.19$~R$_\Earth$) orbiting at $0.163$~au. TOI700-d orbits its host star at the close end of the conservative Habitable Zone \citep[HZ,][]{Kasting93}. Initial studies using climate models \citep{TOI700c} have shown that indeed, TOI700-d has the potential to be habitable taking into account large variety of atmospheric composition possibilities. However, the study also shows that successful measurements of the planetary atmosphere by the James Webb Space Telescope \citep[JWST e.g.,][]{JWST06,Kalirai18} are very unlikely due to the expected signal uncertainty. 

The conservative HZ definition typically refers to the region where the combination of the orbital distance of the planet, and the host star's luminosity defines a planetary surface temperature ranging between $0-100$~C$^\circ$. Thus, such a planet could have water existing on its surface in a liquid form --- an assumed basic requirement for life to form. However, there are many other factors that modify the planetary temperature, such as greenhouse gases, clouds, composition, plate tectonics, and outgassing, which are related to the internal dynamics and the coupling between the planetary surface, atmosphere, and possibly oceans \cite[see e.g., reviews by][]{Lammer2009,Kaltenegger17}. Another aspect influencing the planet habitability is the space environment in its vicinity. This influence can be divided into that of the photon radiation environment, and that of the particle radiation environment. 

Short-orbit exoplanets are expected to receive intense EUV and X-ray radiation \citep{Sanz-Forcada11}, which can lead to higher atmospheric hydrostatic temperature, an inflation of the atmosphere (larger scale-height), and high atmospheric escape rate due to thermal acceleration. Such high escape rates from terrestrial planets were predicted by a number of studies \citep[e.g.,][]{Lammer2003,Penz2008,GarciaSage2017,Johnstone2019}. Extensive work has also been done on escape from gas-giant exoplanets but this is beyond the scope of this paper. The possible influence from Stellar Energetic Particles (SEP) on the HZ of M-dwarf planets has recently been investigated by \cite{Struminsky18} and \cite{Fraschetti19}. These studies have shown that M-dwarf planets may suffer from intense SEP radiation a few orders of magnitude higher than that of an extreme space weather event on the Earth. On the other hand, short planetary orbits are expected to be well shielded from cosmic-ray radiation, especially if the host star has a stronger magnetic field than the Sun \citep[][]{HeliophysicsBook}.

The low-energy particle radiation around short-orbit planets is also expected to be extreme. Specifically, the ambient Stellar Wind (SW) density and the Interplanetary Magnetic Field (IMF) can be 1-3 orders of magnitude higher than the solar wind conditions at $1$~au, leading to much higher SW dynamic pressure \citep[e.g.,][]{Cohen2014,Cohen15,Vidotto15,Garaffo16,Garraffo17}. This extreme SW could erode the planetary atmosphere in the same manner as the Martian atmosphere is stripped by the much weaker solar wind \citep{jakosky2018loss}. Recent studies have shown that the interaction of exoplanetary atmospheres with an extreme SW may indeed lead to high atmospheric loss rate \citep[e.g.,][]{Kislyakova2014,Cohen15,Dong17}. The SW can also lead to atmospheric escape via sputtering \citep[e.g., recent MAVEN observations from Mars,][]{Leblanc18}. On the other hand, recent work by \cite{Vidotto18} suggests that intense SW sputtering on a bare-surface planet could actually lead to a buildup of a neutral atmosphere. 

In the context of planet habitability, 
characterizing the stellar environment near a given planet is crucial for determining whether that planet can sustain its atmosphere over a long time (i.e., compared to the planet's lifetime). There are at least two means of protection for the planetary atmosphere. First, the planet can have a substantial, thick, Venus-like atmosphere. In this case, the atmosphere may be so thick that it can survive erosion and evaporation. However, \cite{Cohen15} have shown that even a Venusian atmosphere may suffer from a high atmospheric loss rate. In addition, one needs to explain how such a thick atmosphere can build up at short orbit in the first place. Second, a strong, internal planetary magnetic field is expected to protect the planetary atmosphere by deflecting the SW. However, in some cases, the planetary field may actually enhance the escape of ions via processes such as the ambipolar electric field and wave-particle interaction \citep{Strangeway2005,Strangeway10a,Strangeway10b,Strangeway:2012,Dong19,Egan19}. In general, we do not have much information about the expected magnetic fields in exoplanets beyond scaling laws such as in \cite{Christensen10}. 

In this paper, we aim to characterize the space environment around the three planets in the TOI700 system, and in particular, the space environment of TOI700-d, which potentially could be habitable. We use state-of-the-art models to obtain the conditions of the stellar corona and the SW near the planets, to capture the interaction of the planetary magnetosphere with the SW, and to estimate the heating of the planetary upper atmosphere as a result of this interaction. We limit this study to an Earth-like, magnetized planet, while a followup investigation of a non-magnetized planet is to follow.

We describe our modeling approach and the different models used here in Section~\ref{Models}, and detail the results of each model in Section~\ref{Results}. We discuss our main findings in the context of the habitability of the TOI700 planets in Section~\ref{Discussion}, and draw our conclusion in Section~\ref{Conclusions}.


\section{Modeling Approach}
\label{Models}

In order to model the space environment of the TOI700 system, we use a set of physics-based models including a model for the stellar corona and stellar wind, a model for the planetary magnetosphere, and a model for the planetary ionosphere. Our modeling domain starts from the stellar chromosphere, where we propagate the solution towards the planets and down to their upper atmosphere. This approach enables us to perform powerful and extensive {\it photosphere-to-atmosphere} investigation of the space environment and its influence on the planet. The details about each of the models, their input/output, and their setting are described in the following sub-sections.

\subsection{Modeling the Stellar Corona and Stellar Wind}
\label{SC}

\subsubsection{The Alfv\'en Wave Solar Model}
\label{AWSOM}

We use the Alfv\'en Wave Solar Model \citep[AWSOM,][]{Vanderholst14} to simulate the stellar corona and SW of TOI700.  AWSOM is a three-dimensional Magnetohydrodynamic (MHD) model that is driven by the distribution of the radial magnetic field at the photosphere (magnetogram) as its boundary condition. For a given set of magnetic field boundary conditions, AWSOM self-consistently calculates the stellar coronal heating and SW acceleration assuming these processes are driven by Alfv\'en waves, while taking into account other thermodynamic processes, such as electron heat conduction and radiative cooling. The model produces a three-dimensional solution for the plasma parameters in the domain between the low corona and interplanetary space (the stellar corona and SW). These solutions can be steady-state solutions for the given magnetic field input (used in this paper), or time-dependent solutions that can capture dynamic changes in the corona \cite[e.g.,][]{Jin17,AlvaradoGomez19}. The model has been extensively validated against solar data, and it has also been widely used to simulate the coronae and winds of Sun-like stars with different spectral types \citep[see e.g.,][for more details]{Cohen15,Garaffo16, AlvaradoGomez16,Garraffo17,Dong18}. 

Our AWSOM simulation uses a spherical grid which is stretched in the radial direction with a radial grid size of $\Delta r=1/20R_\star$, and an angular grid size of $2.8^\circ$ near the inner boundary. We dynamically refine the grid near the astrospheric current sheet in order to better resolve it. The simulation domain extends to $r=100R_\star$ in order to capture the orbit of TOI700-d.

\subsubsection{Proxy Input Data for TOI700}
\label{SC_TOI700}

Magnetogram data are very limited for stars and only available through Zeeman-Doppler Imaging \citep[ZDI,][]{Semel:80}. ZDI measures the polarization (Stokes) parameters from the starlight, which can be related to the transverse and line-of-sight magnetic field. A fitting procedure is applied to these observations to produce latitude-longitude maps of the stellar magnetic field. The method is limited to relatively fast-rotating stars (faster rotation increases the Doppler shift and the spatial resolution of the magnetic maps) and it only provides a large-scale approximation for the stellar magnetic field. 

At the time of writing, there are no ZDI data available for TOI700. Therefore, we follow a similar approach to \cite{Garraffo17} and \cite{AlvaradoGomez19a}, using a proxy star for which ZDI data are available. The star closest in its parameters to TOI700 is CE~Bootis (Gliese 569). Table~\ref{table1} compares the fundamental parameters that are relevant for our model. In general, AWSOM accounts for the stellar radius, mass, and rotation period, in addition to the input magnetic field. The two stars are very close in terms of stellar type, mass, and radius. However, CE Boo's rotation period is much faster than TOI700 ($14.7$~days vs. $54$~days), it is younger, and it is also much more active in the EUV/X-ray range \citep{Rowe07,Donati08}. In order to account for these differences, we follow the relations of  \cite{Vidotto14} and scale the CE Boo magnetogram to 10\% of its original magnitude. Since the magnetogram is roughly dipolar, the dipole field strength of the scaled-down magnetogram is of the order of $20-30$~G \citep[the large-scale complexity for CE Boo's Rossby number is very low][]{Garraffo.etal:18}.

ZDI maps do not include any small-scale magnetic features, which may impact the solution. Therefore, we also use a scaled-up solar magnetogram to provide additional, independent estimation of the SW conditions around TOI700-d. We choose a solar magnetogram for an epoch where the magnetic large-scale topology of the solar field was similar to that of CE Boo. Such a period is Carrington Rotation (CR) 2091, corresponding to  between Dec 7 2009 and January 3 2010 (during the peak of the extended solar minima). We drive the model using a Michelson Doppler Imager \citep[MDI,][]{Scherrer95} magnetogram, which provides high-resolution input to AWSOM (including small active regions). We scale the MDI magnetogram with a scaling factor of 15 to obtain a dipole field strength that is similar to the scaled-down CE Boo map. 

\begin{figure*}[h!]
\centering
\includegraphics[width=7.in]{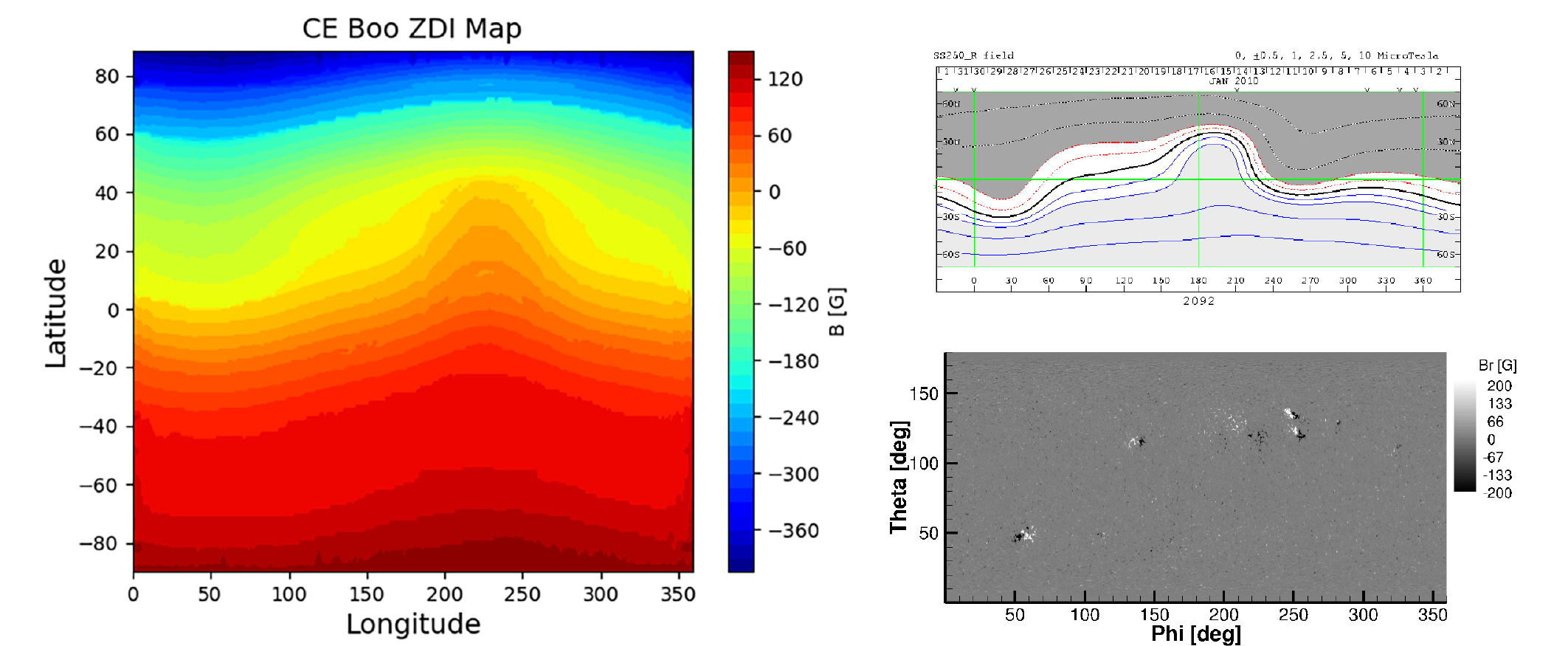}
\caption{Left: ZDI map of CE Boo as reconstructed from \cite{Donati08}. Top right: source surface ($r=2.5R_\odot$) distribution of the Sun's magnetic field during CR2091 obtained by the Wilcox Solar Observatory. Bottom right: photospheric magnetic field during CR2091 obtained by SOHO's Michelson Doppler Imager (MDI). Solar data and figures obtained from  \url{http://sun.stanford.edu/synop}.}
\label{fig1}
\end{figure*}

The combination of a scaled-down stellar proxy and a scaled-up solar proxy provides a reasonable constraint for the input of our modeling of the space environment of TOI700-d. Figure~\ref{fig1} shows the ZDI map of CE Boo, a source surface distribution of the solar field during CR2091, which shows the large-scale topology of the Sun's magnetic field, and the high-resolution MDI magnetic field map. 

\subsection{The Planetary Magnetosphere and Ionosphere}
\label{GMIE}

\subsubsection{A Global Magnetosphere Model}
\label{GM}

AWSOM enables us to extract the plasma parameters along the orbits of the TOI700 planets. The extracted SW conditions along the orbit of TOI700-d are used to drive an MHD model for the planetary magnetosphere. To do so, we convert the coordinate system of these upstream conditions from the star-centric coordinates to the Geocentric Solar Ecliptic \citep[GSE,][]{Kivelson.Russell:95}, assuming the planetary magnetic field is aligned with the vector perpendicular to the stellar system ecliptic plane. 

The Global Magnetosphere (GM) MHD model is a version of the {\it BATSRUS} model \citep{Powell:99,Toth.etal:12}. The model solves the MHD equations for the three-dimensional distribution of the plasma parameters between a spherical inner boundary at $r=2$~R$_p$ and the outer boundary ranging from $100$~R$_p$ on the day side and $r=224$~R$_p$ on the night side. The smallest grid cell size in our GM simulation is $0.3$~R$_p$, where the higher resolution is set around the inner boundary. 

In the simulations here, we assumed an Earth-size magnetized planet, and investigate a range of planetary magnetic fields of $0.3$~G (Earth-like field), $1$~G, and $4$~G (Jupiter-like field). We leave the possibility that TOI700-d is non-magnetized for a future investigation.  

\subsubsection{Ionosphere Model}
\label{IE}

\begin{figure*}[h!]
\centering
\includegraphics[width=6.in]{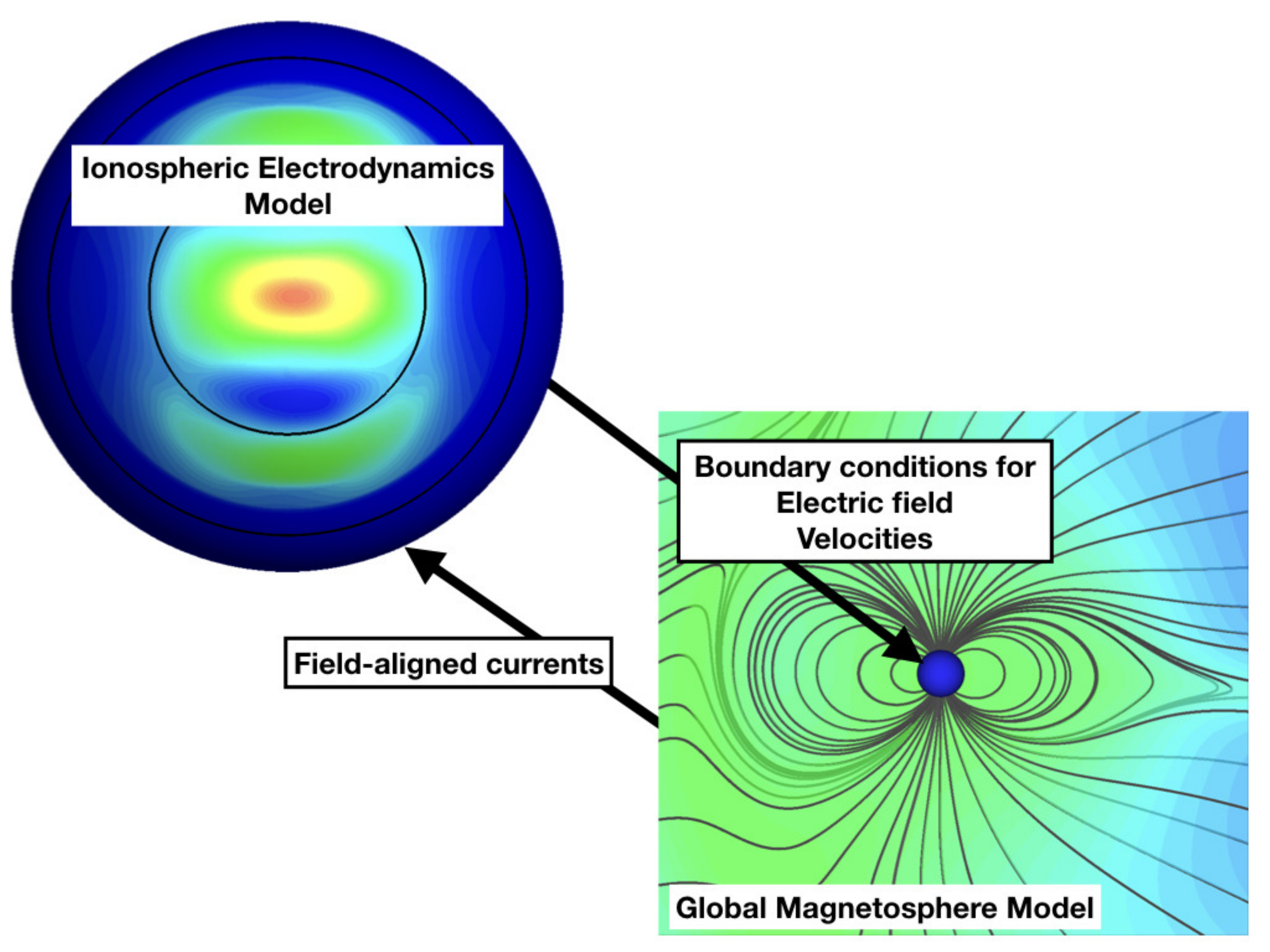}
\caption{GM-IE code coupling procedure is demonstrated using random GM and IE solutions that {\it do not} represent any actual result from this study. The three-dimensional magnetic field lines from GM are used to calculate the field-aligned currents, that are mapped down to the ionosphere assuming a dipole field. The IE model is using these currents to calculate the electric potential, which is then used to set up the inner boundary for the electric field and velocity at the GM domain. For the Earth and TOI700-d, the assumed height of the IE module is $120~km$. Here, the results of the IE module show color contours of the Joule Heating distribution over the Northern hemisphere.}
\label{fig2}
\end{figure*}

In order to capture some of the properties of the upper atmosphere of TOI700-d, we couple the GM model with a model for the planetary ionosphere. The Ionospheric Electrodynamics \citep[IE,][]{Ridley04} model receives the distribution of field-aligned currents from the GM model assuming a dipole field mapping. The field-aligned currents are calculated assuming $\mathbf{J}=\nabla\times\mathbf{B}/\mu_0$ (neglecting the displacement current), with the field aligned current, $j_\parallel=\mathbf{J}\cdot\mathbf{B}$, representing the precipitating electrons moving down from the magnetosphere towards the planetary upper atmosphere. IE uses these currents to calculates the electric potential distribution on a two-dimensional sphere (at a height of $120\;km$ for the case of the Earth). This mapping of the electric potential is then used to self-consistently calculate the electric field and velocity on the inner boundary of the GM model. Figure~\ref{fig2} shows a schematic of the coupling procedure. It is possible that ionospheres of exoplanets are different than that of the Earth (different altitude, thickness, and structure). However, obtaining the detailed ionospheric information would require detailed information about the atmospheric composition, which is unknown, as well as detailed modeling of the planetary upper atmosphere. Thus, we limit ourselves here to an Earth-like ionosphere. 

The main advantage of using the GM-IE coupling here (beside the more physically-constrained solution), is that the IE model provides the ionospheric Joule Heating (JH), which is the result of the current dissipation in the finitely-conducting (i.e., resistive) ionosphere. Such a Joule heating is driven by the incoming SW and it also depends on the ionospheric conductance. The conductance itself depends on the properties of the planetary atmosphere and the stellar radiation. Thus, it is a good measure of how the space environment affects the planetary upper atmosphere. If the SW is intense, a very high JH is expected. We use here a uniform conductance with values of $1$~S and $10$~S,  which represent the range of height-integrated conductances in the Earth's ionosphere \citep[e.g.,][]{Ridley04,Sheng14}.  In reality, the conductance is not uniform and it is affected by the atmospheric composition, ion-neutral interaction, and the stellar EUV radiation. However, this complicated calculation is beyond the scope of the study we present here as it requires the inclusion of a self-consistent, global ionosphere-thermosphere model. Here, we use the simplified current-driven circuit, with $JH=j^2_\parallel/\Sigma_p$, where $\Sigma_p$ is the ionospheric constant, height-integrated conductivity (Pedersen conductivity). We discuss the limitations of this relation further in Section~\ref{Conductances}.


\section{Results}
\label{Results}

All the results we obtain include two simulations of the AWSOM model. One with the stellar ZDI input, and one with the solar MDI input. 

\begin{figure*}[h!]
\centering
\includegraphics[width=3.25in]{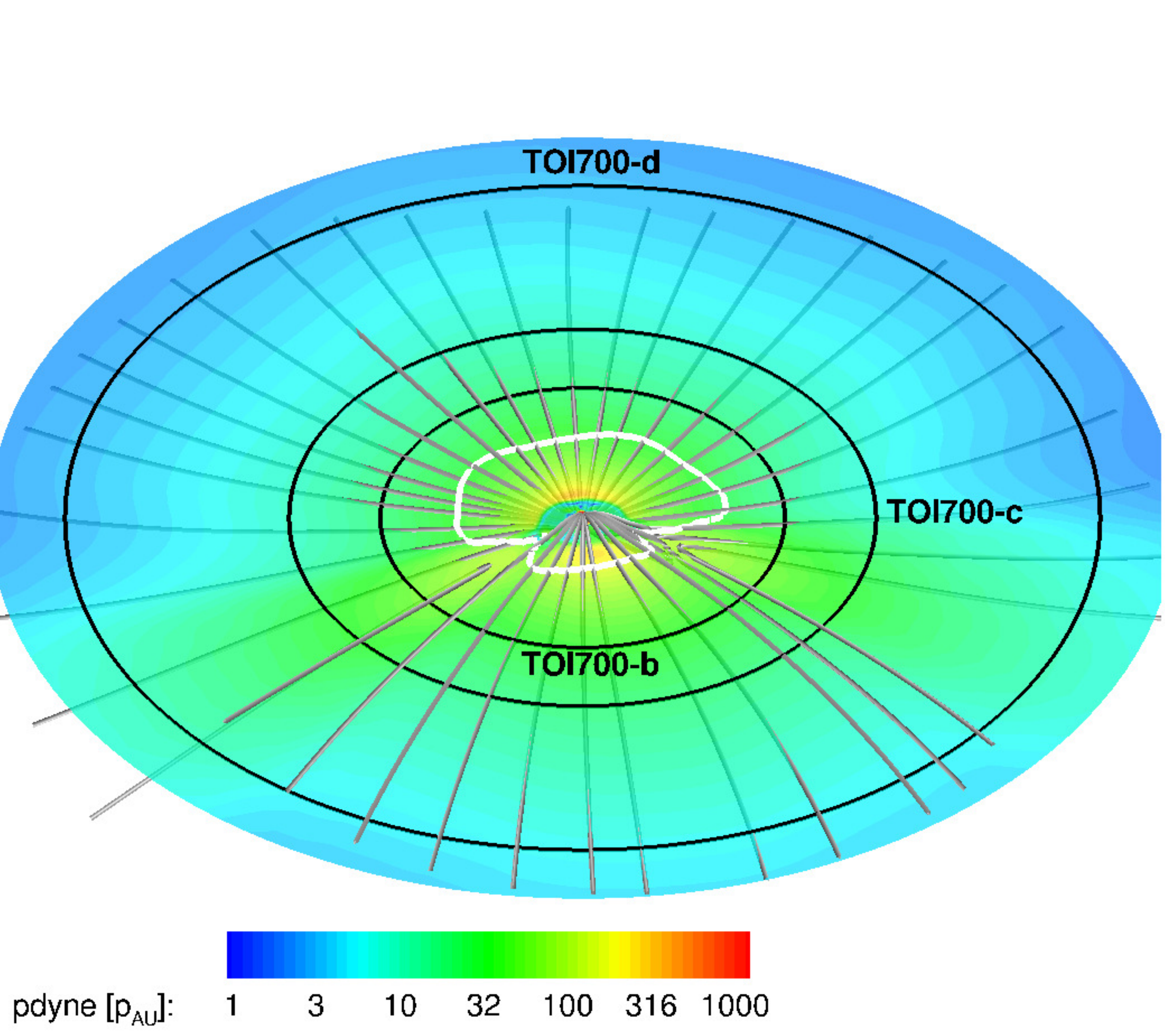}
\includegraphics[width=3.25in]{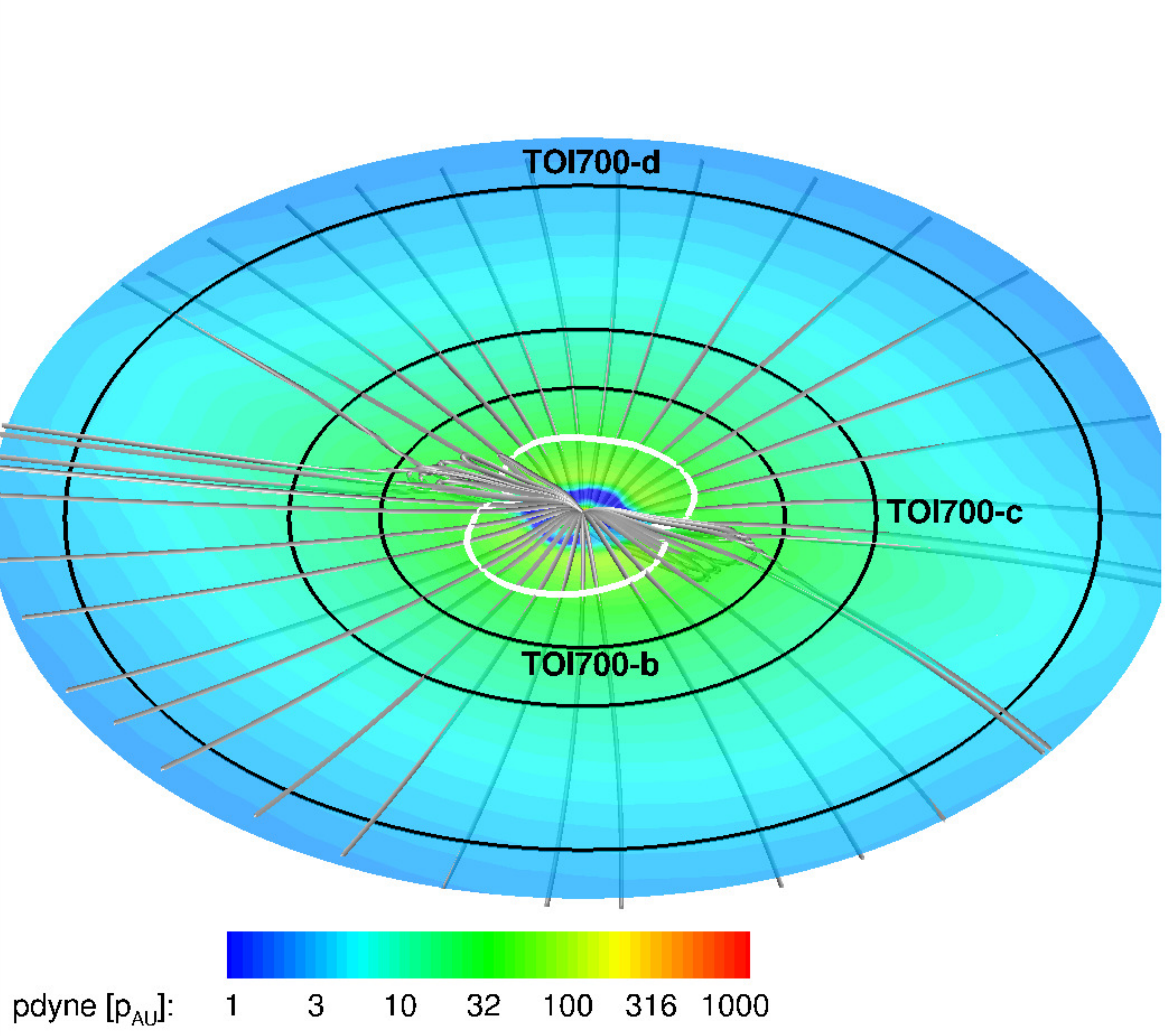}
\caption{AWSOM solutions for the scaled-down CE Boo input (left) and the scaled-up solar input (right). Plots show the equatorial plane colored with contours of the ratio of the SW dynamic pressure to the typical dynamic pressure at 1~AU. Black circles show the orbits of the three TOI700 planets, the white solid line represents the Alfv\'en surface, and gray lines show selected magnetic field lines.}
\label{fig3}
\end{figure*}

Figure~\ref{fig3} illustrates the stellar wind solution using AWSOM. It shows the equatorial plane colored with contours for the ratio between the dynamic pressure in the simulation to the dynamic pressure of the solar wind with typical solar wind parameters at $1$~au [$P_{d_{AU}}$]. These typical parameters include solar wind number density, $n_{sw}=5$~cm$^{-3}$, and solar wind velocity, $v_{sw}=500$~km~s$^{-1}$. Also shown in the figure are the orbits of TOI700-b, TOI700-c, and TOI700-d, together with the Alfv\'en surface location, and selected coronal magnetic field lines. The expected SW dynamic pressure near TOI700-d ranges between $5-10~P_{d_{AU}}$ for the solar proxy, and $5-50\;P_{d_{AU}}$ for the stellar proxy.  

\begin{figure*}[h!]
\centering
\includegraphics[width=6.in]{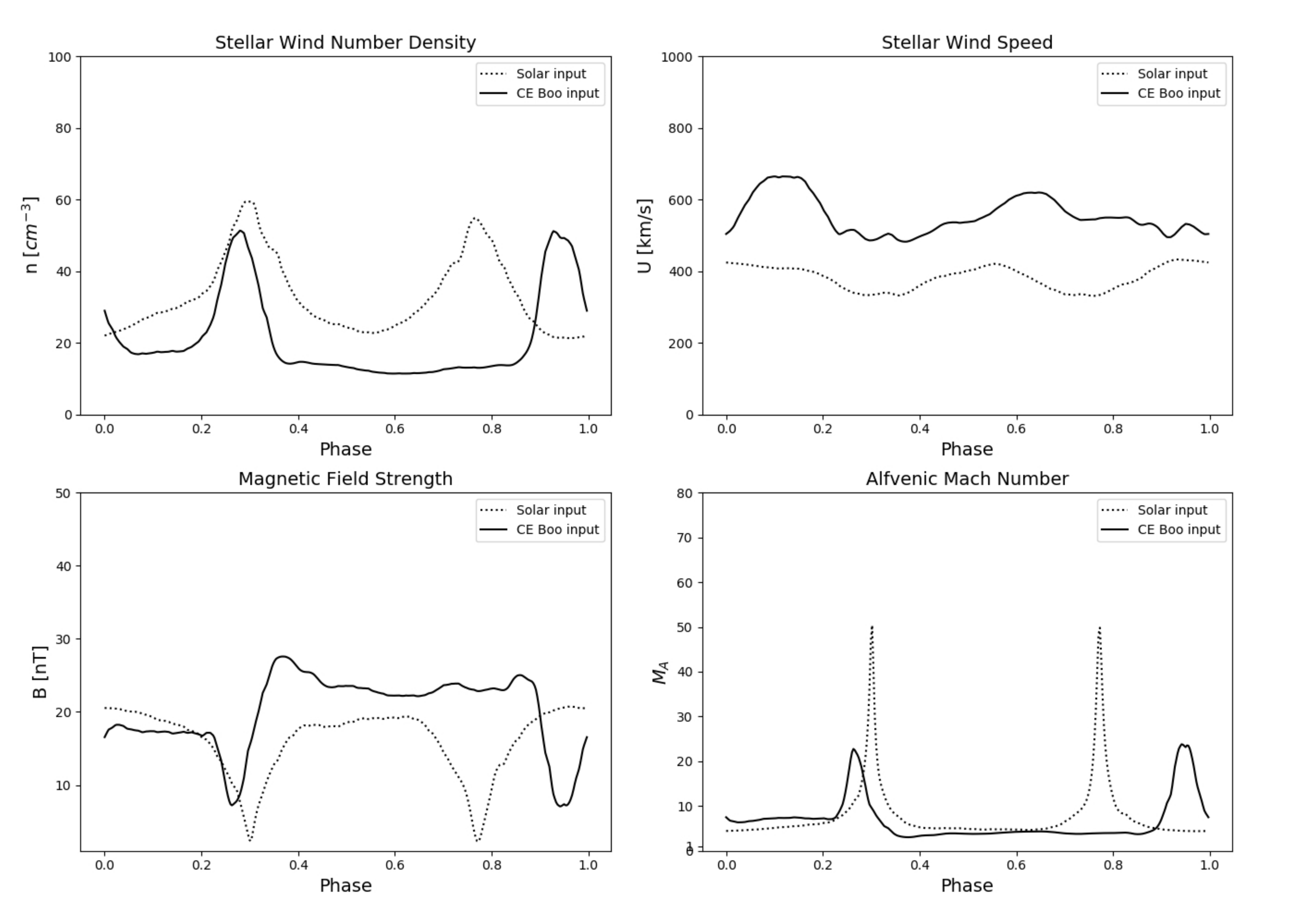}
\caption{SW conditions as extracted along the orbit of TOI700-d shown as a function of the orbital phase. Parameters are SW number density (top left), SW speed (top right), SW magnetic field strength (bottom left), and SW Mach number. Results are for the CE Boo stellar input (solid line) and for the solar input (dashed line).}
\label{fig4}
\end{figure*}

Figure~\ref{fig4} shows the extracted SW parameters along the orbit of TOI700-d for the two proxy solutions. The order of magnitude of the SW parameters for the two independent proxies is similar, and both proxies show two crossings of the stellar helmet streamers along the orbit, where the density increases. The two transitions also suggest a crossing of the astrospheric current sheet along the orbit, indicated by a drop of the magnetic field strength. Overall, both the density and magnetic field strength are about 5-10 times higher than their typical ambient solar wind values. The SW speed ranges between $400-650$~km~s$^{-1}$. This means that the SW origin around the orbit of TOI700-d is mostly slow SW \citep{McComas07}, which is not surprising for a nearly straight dipolar stellar field topology and a planetary orbit confined close to the ecliptic plane. It is possible that TOI700-d experiences high SW streams during periods of stellar magnetic field reversals (if such reversals exist), or when the orbit is facing an equatorial coronal hole, in a similar manner to the Earth. The Alfv\'enic Mach number, $M_A=u_{sw}/v_A$, is given by the ratio of the SW speed, $u_{sw}$, to the local Alfv\'en speed, where $v_A=B_{sw}/\sqrt{4\pi \rho_{sw}}$, and $B_{sw}$ and $\rho_{sw}$ are the SW magnetic field strength and mass density, respectively. The SW along the orbit of TOI700-d is found to be always super Alfv\'enic ($M_A>1$), where $M_A$ can be as high as 20-40 near the helmet streamer crossings, resulting in a very strong MHD shock in front of the planetary magnetosphere. 

\begin{figure*}[h!]
\centering
\includegraphics[width=7.in]{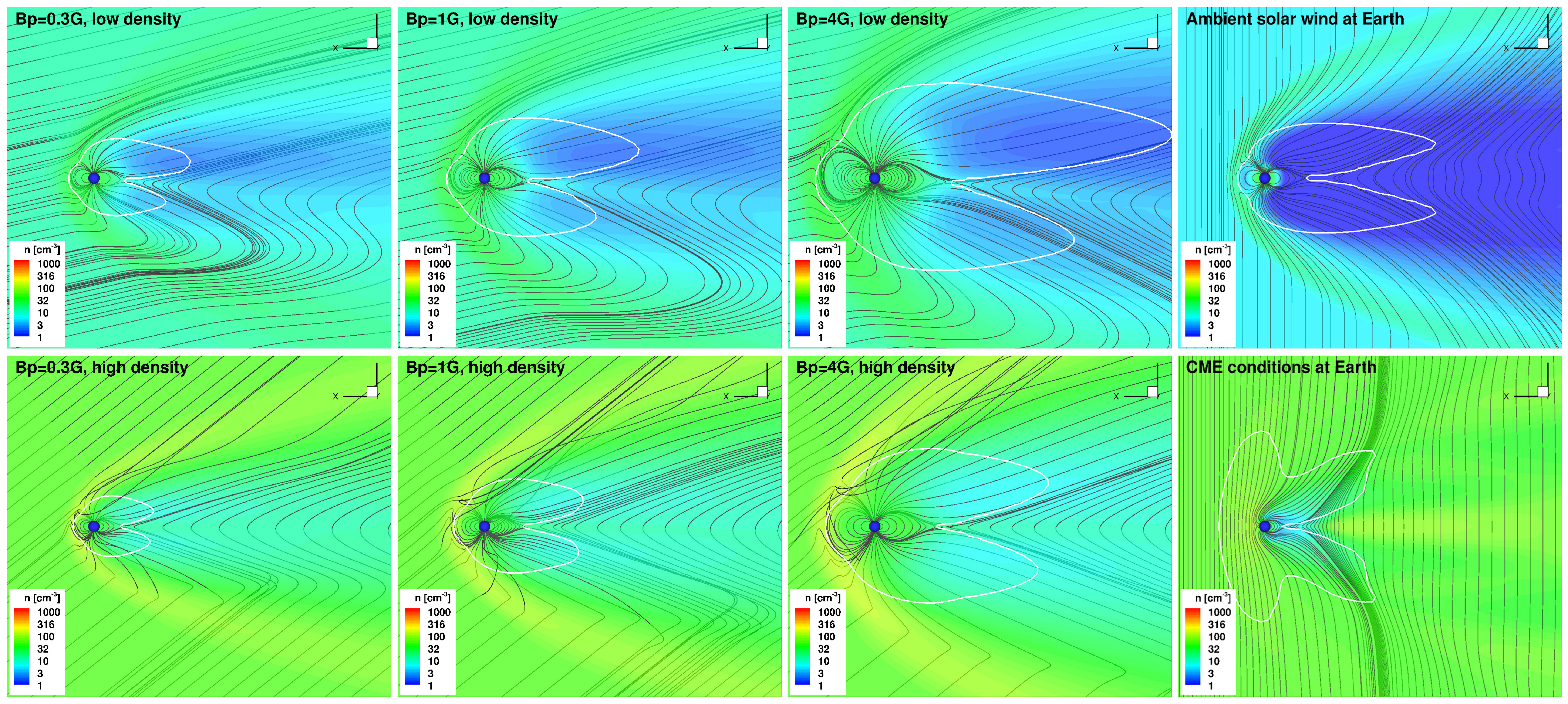}\\
\includegraphics[width=7.in]{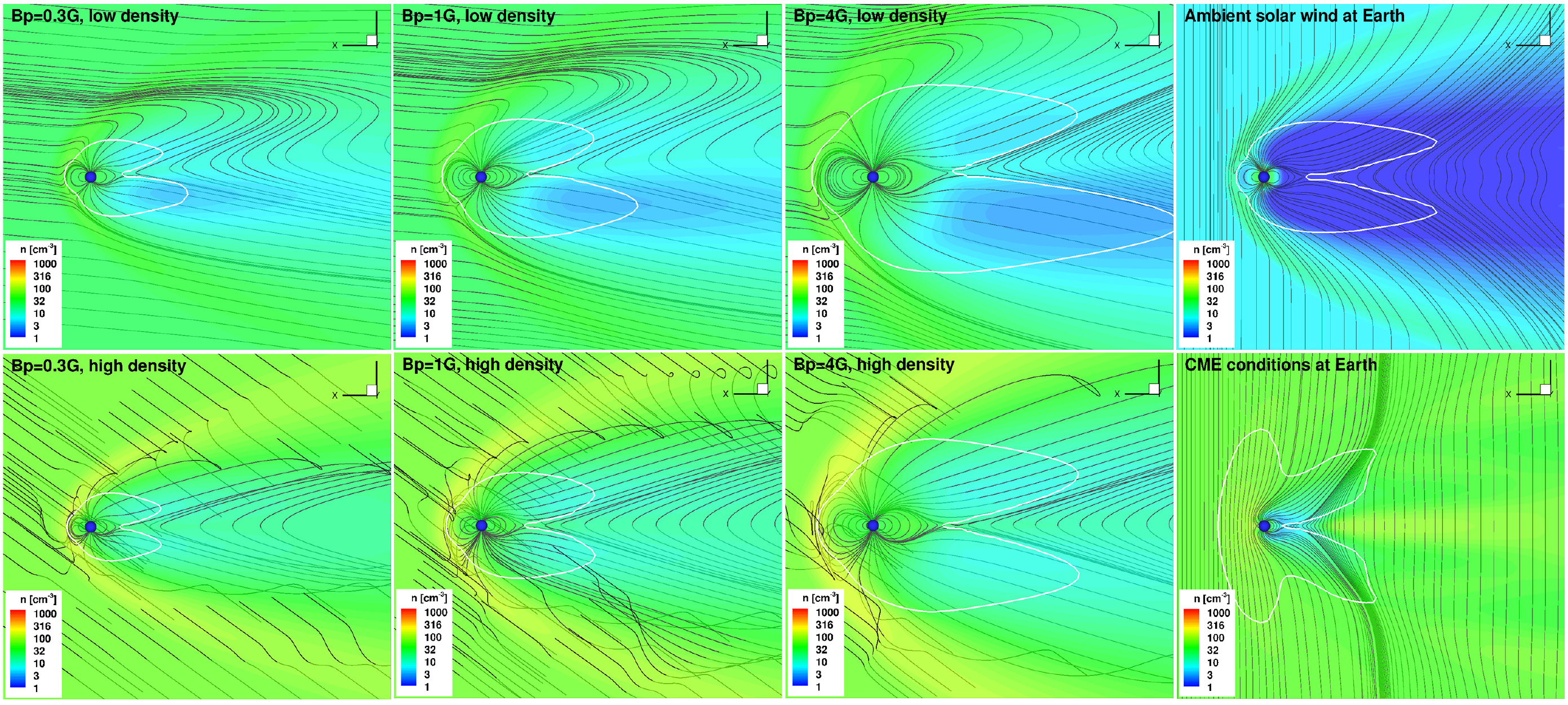}
\caption{Magnetospheric structure displayed on a day-night meridional plane for the CE Boo input (top two rows) and for the solar input (bottom two rows). For each set of two rows, the top row shows the results for the lowest SW density point, and the lower row shows results for the highest SW density point. Planes are colored with number density contours, where the Alfv\'en surface is shown as a solid white line, and some selected magnetic field lines are shown in grey. The first three left columns show results for planetary magnetic field of $0.3G$, $1G$, and $4G$, respectively. The right column shows the results using ambient solar wind conditions (top) and a CME event (bottom).}
\label{fig5}
\end{figure*}

In Figure~\ref{fig5}, we show the predicted structure of the magnetosphere of TOI700-d for the two proxy inputs and for the three planetary magnetic field strength values. For reference, we also show the magnetospheric structure using SW input representing typical ambient solar wind conditions, and solar wind conditions during a strong Coronal Mass Ejection (CME) event. In both cases we choose a SW field that is aligned opposite to the planetary field so that the JH is maximized: the SW magnetic field is positive or northward, pointing in the direction opposite to the planetary, positive dipole field. This is opposite to the Earth's case, where a  southward SW field is opposite to the Earth's present-day, negative dipole field. These SW input datasets are summarized in Table~\ref{table2}, where we use the same model setting for these additional simulations. Each plot shows a meridional cut, which extends from the day to the night side, and is colored with number density contours. Some selected magnetic field lines are shown as black lines. The magnetospheric size is approximated by the surface of $M_A=1$, which is marked as a solid white line. Overall, the magnetospheric structure is similar to that of the Earth, but slightly more compressed on the day side due to the higher SW density for TOI700-d. Clearly, the increase in the planetary field strength from $0.3$~G to $4$~G increases the magnetosphere size (and the magnetosphere standoff distance), as the planetary magnetic field pressure overcomes the SW dynamic pressure. It is worth mentioning that the magnetosphere standoff distance could be obtained from an analytical pressure balance formula \citep[e.g.,][]{Gombosi:04} without the need of the GM simulations. The advantage of the GM simulations is that they provide a better sense of the three-dimensional magnetospheric structure, and they are also needed to obtain the IE results for the ionospheric JH.

\begin{figure*}[h!]
\centering
\includegraphics[width=3.25in]{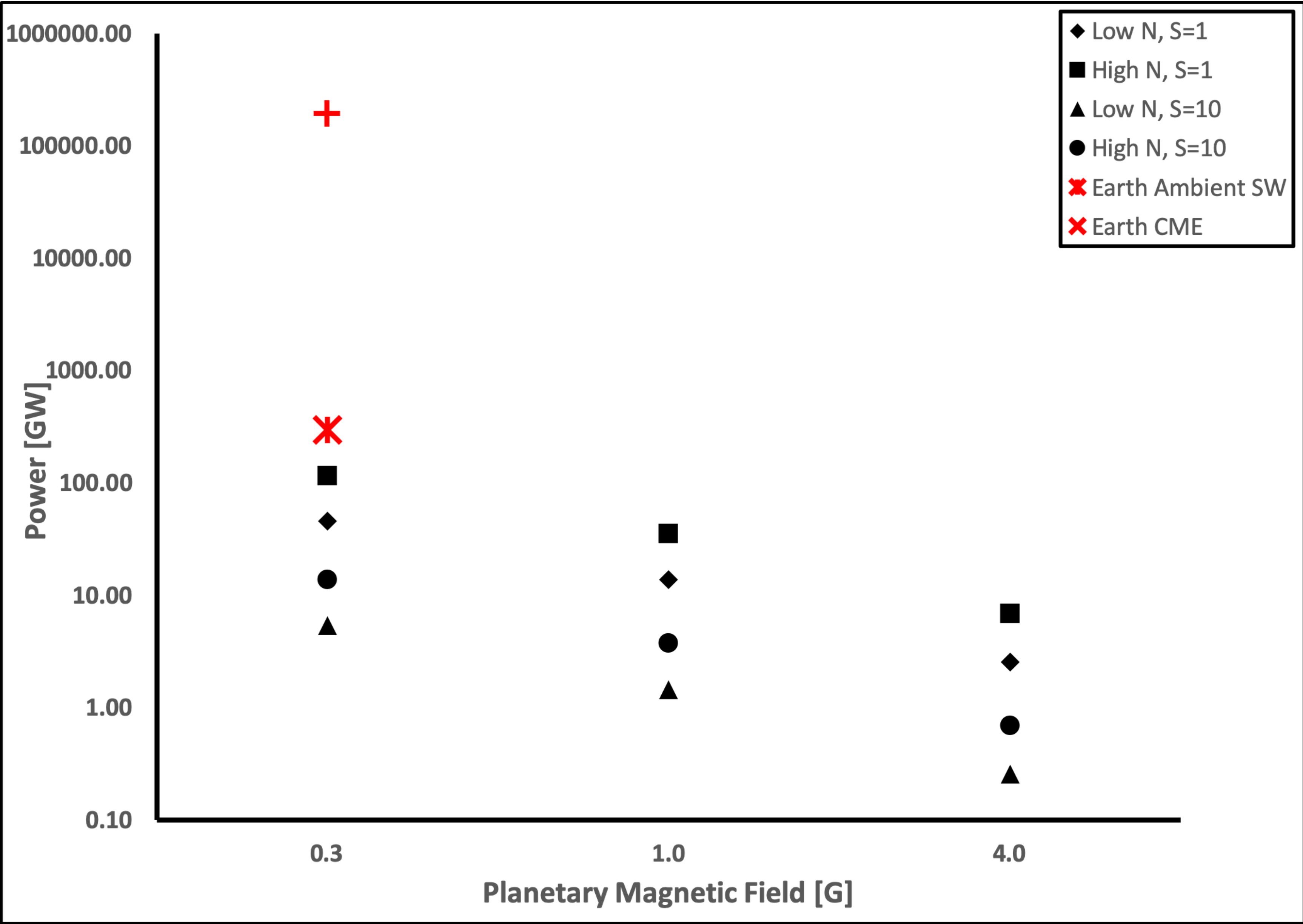}
\includegraphics[width=3.25in]{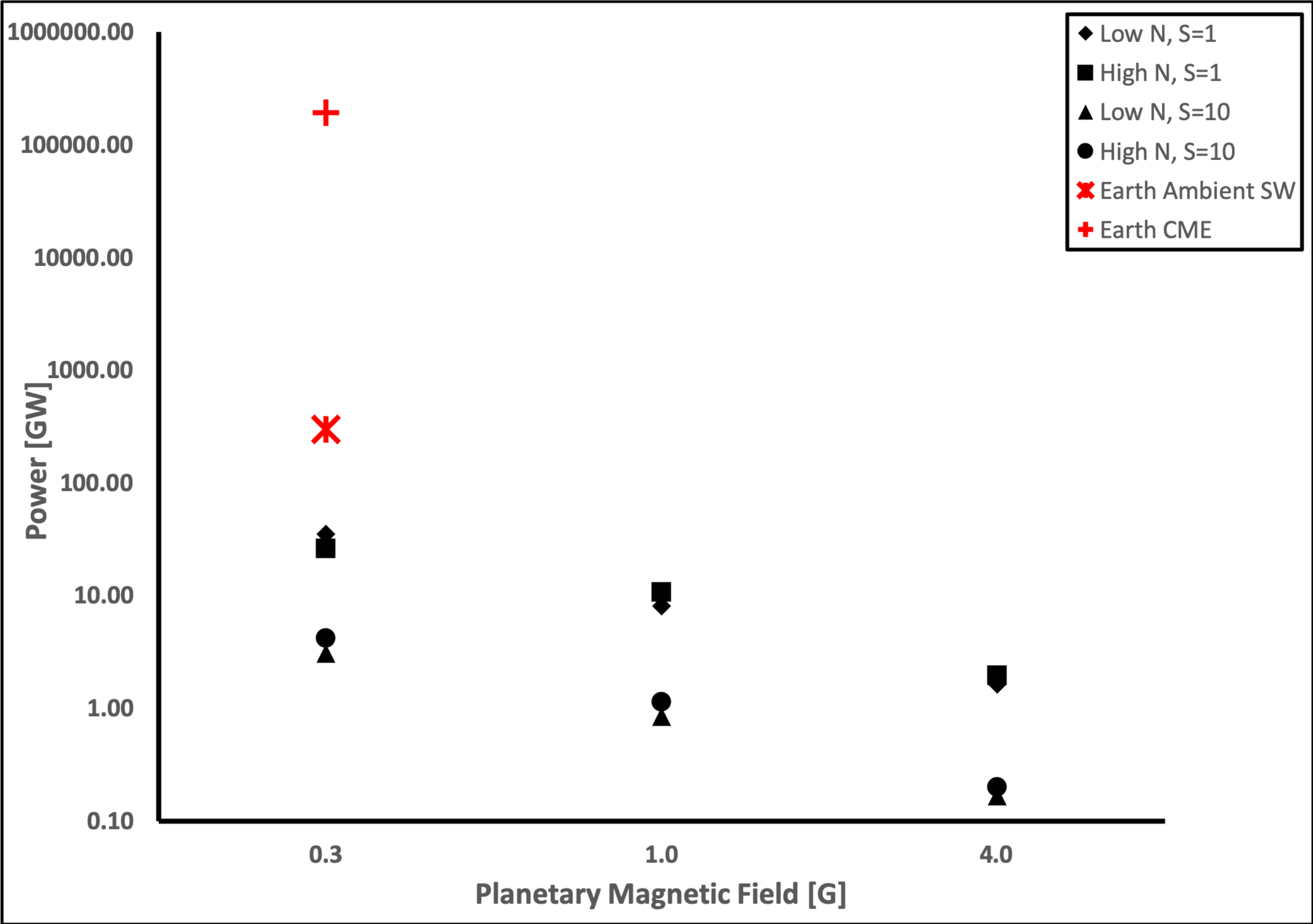}
\caption{JH power as a function of planetary magnetic field strength. Results correspond to the cases shown in Figure~\ref{fig5}, and for ionospheric conductance of $1S$ and $10S$. Also shown in red are JH power for the ambient solar wind at 1AU, and a CME event on Earth.}
\label{fig6}
\end{figure*}

The JH results are shown in units of GWatts (GW) in Figure~\ref{fig6}. The JH associated with the two reference cases for the Earth are also shown. The IE JH solution for each case is given in [mW~m$^2$] so we convert each value to [W~m$^2$] and integrate it over a sphere with an Earth radius (neglecting the ionospheric height of $120$~km) to obtain the total power. The highest JH is obtained for the weak planetary field case of $B_{p}=0.3$~G. For this case, the JH is 10-100 times higher than that of ambient solar wind conditions at Earth, but smaller than the JH during a strong geomagnetic storm. As expected, the JH is weaker for larger values of the ionospheric conductance, and it is smaller for stronger planetary field strength. Interestingly, the JH is higher for higher SW density in the case of the stellar proxy driver. However, the JH is very similar for different densities for the solar proxy driver. This is due to the fact that the SW density differences between the minimum and maximum orbital points are smaller in the solar input, comparing to those of the stellar input.


\section{Discussion}
\label{Discussion}

\subsection{Stellar Wind Impact on the atmosphere of TOI700-d}

The results of our simulations suggest that the physical conditions of the space environment around TOI700-d are not that extreme at the present time, and they are actually less extreme than the solar wind conditions during a major solar storm. Nevertheless, the SW conditions are stronger than the conditions of the ambient solar wind near the Earth (higher dynamic pressure). Therefore, the long-term, integrated effect of these more extreme conditions may be significant. 

\begin{figure*}[h!]
\centering
\includegraphics[width=3.25in]{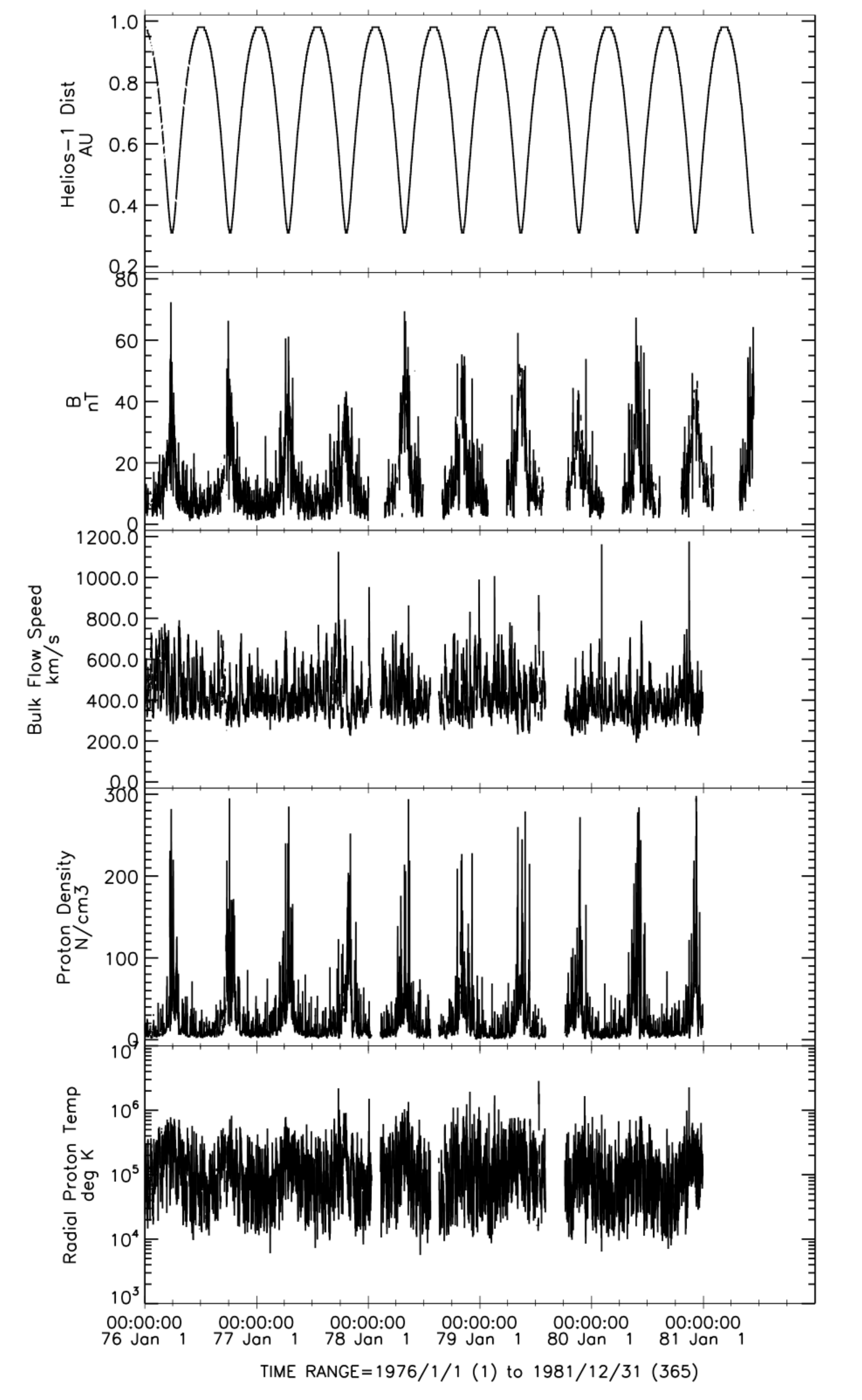}
\includegraphics[width=3.25in]{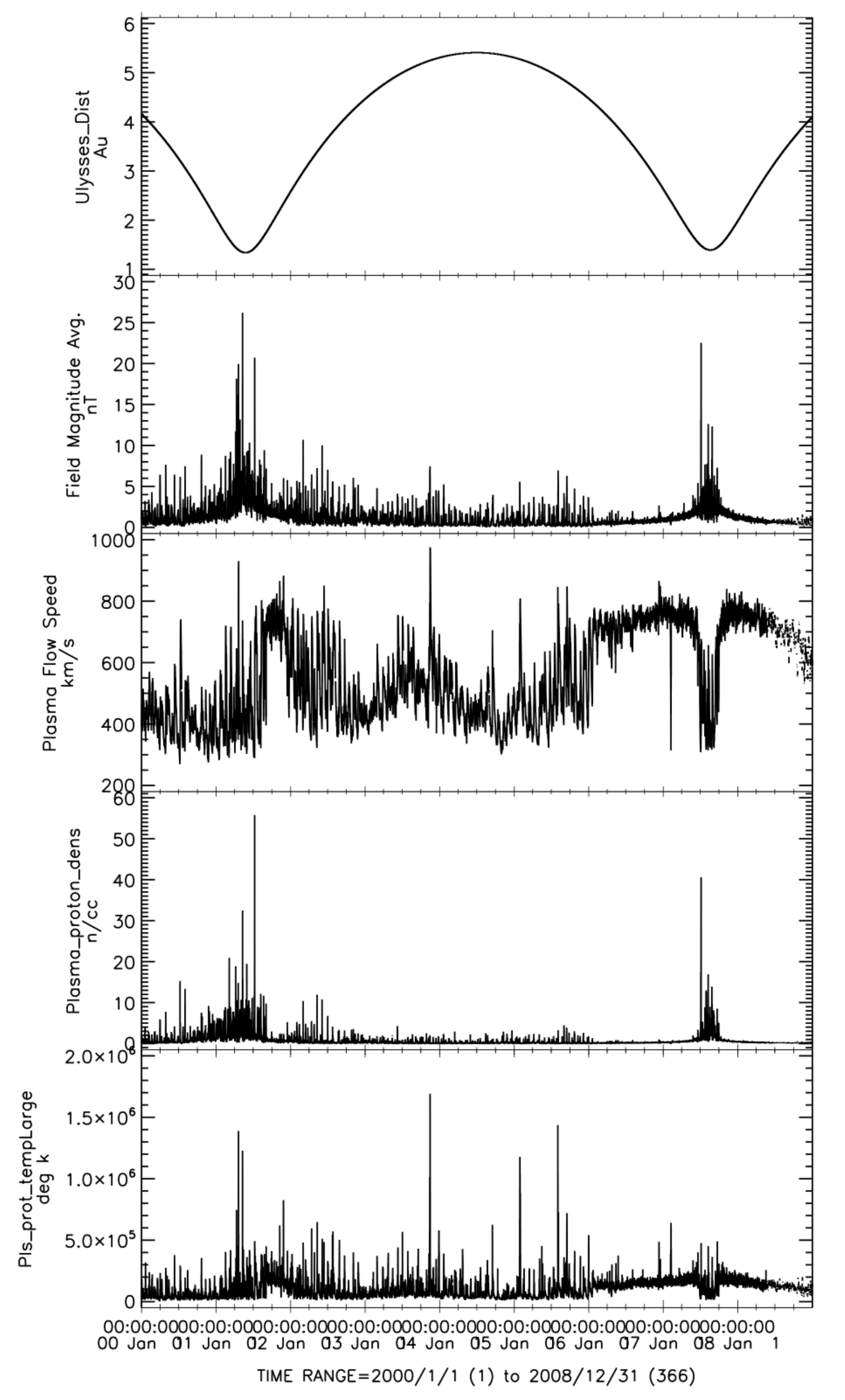}
\caption{Time series of solar wind {\it in situ} measurements as obtained by the Helios mission (left) and the Ulysses mission (right). Rows show (top to bottom) the heliospheric distance from the Sun, magnetic field strength, solar wind speed,  proton density, and proton temperature.}
\label{fig7}
\end{figure*}

As a reference from our own solar system, Figure~\ref{fig7} shows solar wind data from two past missions, obtained from NASA's CDAWEB in-situ data repository at \url{https://cdaweb.gsfc.nasa.gov}. The first data set comprises long-term, 1-hour cadence solar wind conditions as obtained by the Helios spacecraft \citep{Helios} over the course of 5 years at heliospheric distances smaller than $1$~au. The innermost location is at about $0.3$~au, so the Helios data provides information about the solar wind conditions around Mercury's orbit. The other data set comprises long-term, 1-hour resolution solar wind conditions as obtained by the Ulysses spacecraft \citep{Ulysses} over the course of 9 years at heliospheric distances greater than 1AU. The Ulysses data provide some information about the solar wind conditions around Mars' orbit (and beyond). It is important to note that Ulysses had a polar orbit. We use Ulysses data as reference focusing on periods where it measured slow, more dense solar wind (even though the fast solar wind have even lower density).

The simulated parameters of TOI700-d contain density ranging between $40-60$~cm$^{-3}$, magnetic field strength ranging between $10-30$~nT, and wind speed ranging between $400-650$~km~s$^{-1}$. In our discussion here, we do not account for the SW temperature. During its closest encounters ($0.3-0.4$~au near Mercury's orbit), Helios observed magnetic field strength of similar magnitude range. However, the density at these locations is higher than that of TOI700-d---about $100$~cm$^{-3}$ or more. Mercury is a bare planet with no atmosphere. It is believed that the atmosphere has been stripped by the intense solar wind at its orbit  \citep[see, e.g. review by][]{Domingue07}. The intense solar wind has been observed to even interact directly with the planetary surface. This complete loss of Mercury's atmosphere occurred despite its internal magnetic field. The density conditions near TOI700-d seem to be less extreme than those at Mercury's orbit. Thus, compared with Mercury, it is more likely that TOI700-d can sustain its atmosphere. It is clear that the solar wind conditions at Mars' orbit are far less extreme than those at Mercury's orbit, as well as TOI700-d. Nevertheless, there is strong evidence from the MAVEN mission that Mars continuously loses its atmosphere to the solar wind, despite of the solar wind's weak parameters \citep{Leblanc18}. 

Unlike other, previously discovered exoplanetary systems, such as Proxima Centauri b \citep{Garaffo16,Dong17,GarciaSage2017} and Trappist-1 \citep{Garraffo17,Cohen2018}, where SW conditions are more likely to have been too extreme for too long for the planetary atmosphere to survive, it seems more feasible that TOI700-d can sustain an atmosphere. Another factor that may support this is the fact that TOI700-d is an Earth-size planet, larger than both Mercury and Mars, and with a stronger gravitational field to keep the atmosphere in place. Other factors that can affect atmospheric retention, such as the atmospheric composition, and internal magnetic field strength are unknown. 

\begin{figure*}[h!]
\centering
\includegraphics[width=5.5in]{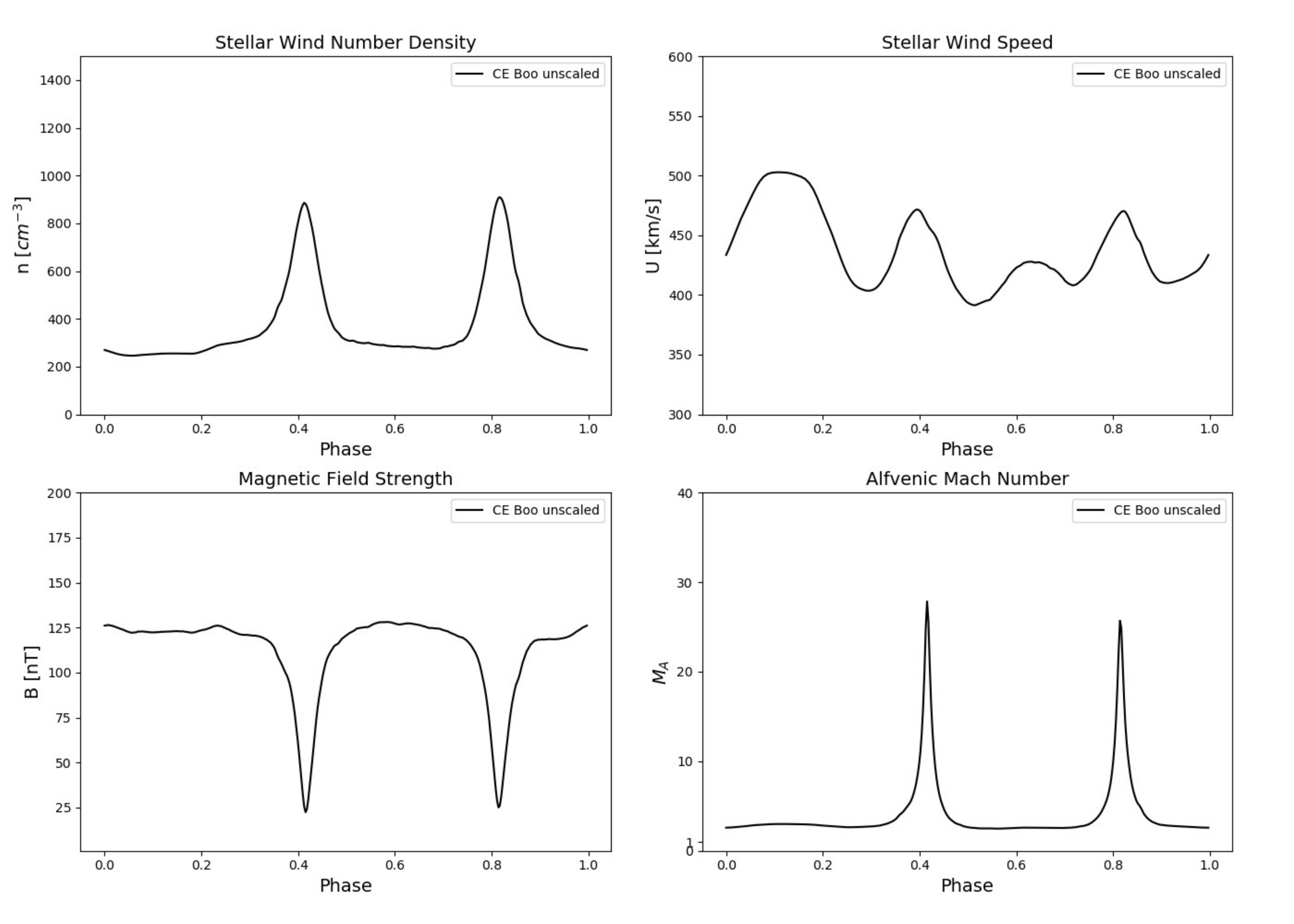}\\
\includegraphics[width=5.in]{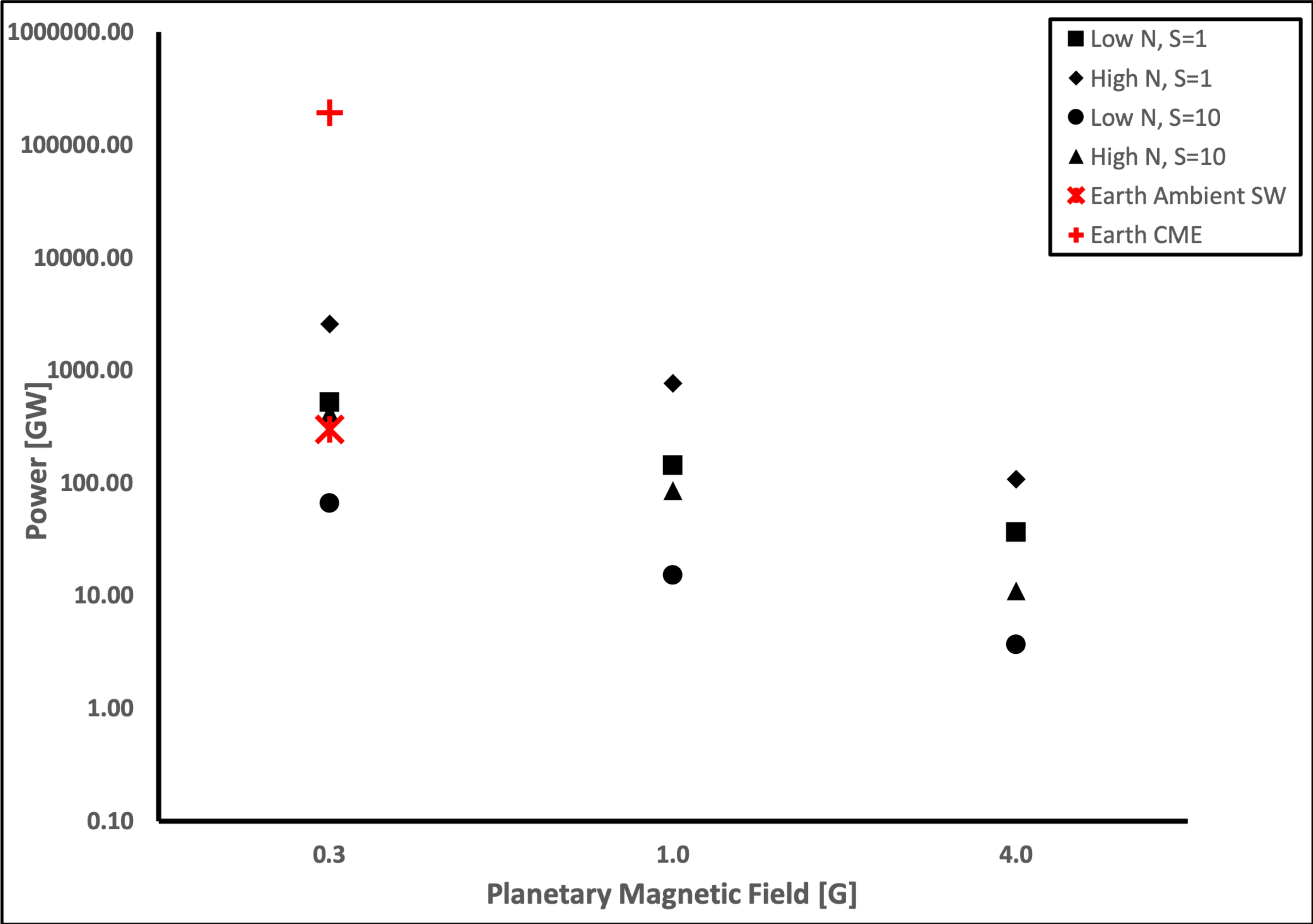}
\caption{SW parameters along the orbit of TOI700-d using unscaled ZDI data of CE Boo (top), and the expected JH power (bottom, similar to Figure~\ref{fig6}).}
\label{fig8}
\end{figure*}

Our results suggest an optimistic scenario for the habitability of TOI700-d. Nevertheless, we stress that the stellar magnetic field plays a major role in defining its ambient SW conditions. Therefore, more data on the stellar magnetic field are required in order to better constrain the habitability of the planet. To set an upper limit for the SW conditions, Figure~\ref{fig8} shows the predicted  conditions along the orbit of TOI700-d and the resulting JH obtained with an unscaled ZDI magnetogram (scaling factor of 1). The SW conditions and JH with this unscaled stellar input are much stronger than the scaled down input data. Even with a scaling factor of 0.2 or 0.3, it is possible that the SW conditions may become sub-Alfv\'enic (due to the stronger stellar field), and the SW density and dynamic pressure will become significantly stronger, leading to a much higher atmospheric loss rate. 

\subsection{The Need for more Self-consistent Calculations of the Atmospheres of Exoplanets}
\label{Conductances}

To demonstrate the complexity of real planetary atmospheres, we use the relation between JH and atmospheric conductivity. In this paper, we use the current-driven circuit relation, $JH=j^2/\Sigma_p$, which assumes simplified, uniform, constant height-integrated conductivity. $j$ in this case is the field-aligned current. In reality, the relations between the SW driver, the currents flowing through the ionosphere, the background planetary magnetic field, and the atmospheric medium, which is composed of ions, electrons, and neutrals are much more complicated and non-trivial. Specifically, here we assume a scalar conductivity. The conductivity can also be different with respect to the direction of the magnetic and electric field vectors, allowing currents to flow differently in each direction. As a result, the dissipation of the currents and the resulting JH may be different in each direction \citep[see e.g.,][]{Kivelson.Russell:95,Gombosi:04}. Moreover, the ionospheric conductance depends on non-trivial relations between the densities of the ions, electrons, and neutrals, as well as the collision frequencies between them. These parameters can drastically change when the planetary atmosphere is highly irradiated by stellar ionizing radiation. Thus, obtaining realistic conductivities and their spacial distributions is a highly challenging task.

If we consider a more complicated conductivity pattern, the relation between JH and $\Sigma_p$ can be represented by a voltage generator, which takes the form of $JH=\Sigma_p E^2$. Here, $E=vB$ is the motional electric field, with $v$ being the plasma relative velocity with respect to the neutral atmosphere, and $B$ is the ionospheric magnetic field. The first notable thing is that in this process, $JH$ and $\Sigma_p$ are not inversely proportional anymore. Additionally, we can write $JH=\Sigma_p v^2B^2$. Assuming that $\Sigma_p \propto n e/B$, with $n$ being the ionospheric number density, and $e$ being the electron charge, we note that $JH\propto v^2B$, and it increases for larger planetary magnetic field, similarly to the relation we obtain assuming a  current-driven circuit with $JH\propto j^2 B$. In our simulations here, we did not account for the relation between $\Sigma_p$ and $B$, nor did we account for the ionospheric density. While a more realistic numerical model that accounts for these complex phenomena would be ideal, the simplified approach adopted here provides an initial step towards estimating JH in exoplanets.

The relation between the JH and the atmospheric escape of ions can also be complicated. \cite{Strangeway:2012}, among others, argued that Pedersen currents exist because of the need to impose plasma flows on the ionosphere when the planetary magnetic field lines are connected to the IMF. Moreover, \cite{Strangeway2005} noted that ionospheric JH through large-scale currents preferentially heats ions, increasing the ionospheric scale height, but such an increase in scale height cannot explain the observed escape of ions from the Earth. Thus, an additional mechanism, such as heating due to wave-particle interaction, is needed to explain the further acceleration of these ions. Finally, \cite{Strangeway10a,Strangeway10b} pointed out that, based on solar system observations, ion escape from magnetized and non-magnetized planets seems to be comparable.

In order to estimate the detailed, self-consistent interaction between all parameters that defined the upper atmospheres of exoplanets, and the driving of atmospheric escape (both neutrals and ions) it is necessary to develop three-dimensional, self consistent MHD models for the upper atmosphere, such as the Global Ionosphere-Thermosphere Model \citep{RidleyGITM:2006}. The modeling should be applied to both magnetized and non-magnetized planets. This will enable much better estimates of the global heating of the upper atmosphere, and the resulting mass-loss rates. However, such a model may need data inputs concerning the atmospheric parameters, which might not be available in the near future.


\section{Conclusions}
\label{Conclusions}

We present here a first step in investigating the SW conditions and the heating of the upper atmosphere of TOI700-d using a suite of numerical models that cover the domain between the stellar corona and the upper planetary atmosphere. Due to the lack of information about the stellar magnetic field, we use a scaled-down stellar proxy and a scaled-up solar proxy. We find that the SW conditions and the atmospheric JH at TOI700-d are moderately higher than those of the ambient solar wind around the Earth, and are lower than those of a strong geomagnetic event. Therefore, our models show that TOI700-d stands a good chance to retain its atmosphere {\it under the assumed input and model parameters}. The parameters that are included in our models could be modified, resulting in a different solutions. These include the assumed planetary magnetic field and its orientation, and the assumed stellar magnetic field, which can also be chosen to represent different epochs in the evolution of the planetary and stellar system of TOI700. We are cautious due to the uncertainty in the stellar proxy data, and the need for a more detailed atmospheric modeling, which we plan to perform in our next study. 


\acknowledgments

We thank an unknown referee for her/his useful comments and suggestions. The work presented here was funded by NASA's NExSS grant NNX15AE05G and NASA's Living with a Star grants NNX16AC11G and. Simulation results were obtained using the Space Weather Modeling Framework, developed by the Center for Space Environment Modeling, at the University of Michigan with funding support from NASA ESS, NASA ESTO-CT, NSF KDI, and DoD MURI. Simulations were performed on the Massachusetts Green High Performance Computing Center (MGHPCC) cluster.




\begin{table*}[h!]
\caption {The stellar parameters of CE Boo and TOI700} 
\begin{center}
\begin{tabular}{  p{1in}  p{1.5in} p{0.5in} }
\hline
{\bf Parameter}&{\bf CE Boo}\footnote{\cite{Donati08}},\footnote{\cite{Rowe07}} &{\bf TOI700}\footnote{\cite{TOI700a}} \\
\hline
Spectral Type&M2.5&M2\\
$R_\star~[R_\odot]$&0.43&0.42\\
$M_\star~[M_\odot]$&0.38&0.416\\
$P_{rot}~days$&14.7&54\\
Age [Gyr]&0.1-0.5& $>1.5$\\
\hline
\end{tabular}
\end{center}
\label{table1}
\end{table*}

\begin{table*}[h!]
\caption {Reference Earth SW Parameters} 
\begin{center}
\begin{tabular}{  p{1in}  p{1.35in} p{1.35in} }
\hline
{\bf Parameter}&{\bf Ambient Solar Wind}&{\bf Strong CME Event} \\
\hline
$n\;[cm^{-3}]$&5&50\\
$u\;[km\;s^{-1}]$&500&1500\\
$B\;[nT]$&10&400\\
\hline
\end{tabular}
\end{center}
\label{table2}
\end{table*}


\end{document}